\documentclass[11pt]{amsart}
\usepackage{a4,amssymb,graphicx}

\begin{document}
\renewcommand{\baselinestretch}{1.12}

\markboth{M.~Baake and U.~Grimm}{Diffraction of limit periodic point sets}

\title{Diffraction of limit periodic point sets}

\author{Michael Baake and Uwe Grimm}
\address{Fakult\"{a}t f\"{u}r Mathematik, 
  Universit\"{a}t Bielefeld, 
  Postfach 100131, 33501 Bielefeld, Germany (MB)}
\address{Department of Mathematics and Statistics, 
  The Open University, 
  Walton Hall, Milton Keynes MK7 6AA, UK (UG)}
\address{School of Mathematics and Physics, 
  University of Tasmania, Private Bag 37,
  Hobart, Tasmania 7001, Australia (MB and UG)}

\begin{abstract}
  Limit periodic point sets are aperiodic structures with pure point
  diffraction supported on a countably, but not finitely generated
  Fourier module that is based on a lattice and certain integer
  multiples of it.  Examples are cut and project sets with $p$-adic
  internal spaces. We illustrate this by explicit results for the
  diffraction measures of two examples with 2-adic internal spaces.
  The first and well-known example is the period doubling sequence in
  one dimension, which is based on the period doubling substitution
  rule. The second example is a weighted planar point set that is
  derived from the classic chair tiling in the plane. It can be
  described as a fixed point of a block substitution rule.
\end{abstract} 

\maketitle
\thispagestyle{empty}


\section{Introduction}

The diffraction measure is a characteristic property of a translation
bounded measure $\omega$ on Euclidean space (or on any locally compact
Abelian group) and has important applications in crystallography
because it describes the outcome of kinematic diffraction (from X-rays
or neutron scattering, say). It is the Fourier transform
$\widehat{\gamma}$ of the autocorrelation measure $\gamma$ of
$\omega$. Following the discovery of quasicrystals (which are
generally non-periodic but nevertheless show pure point or Bragg
diffraction), measures with pure point diffraction have been
investigated thoroughly. In parallel, there has also been progress in
understanding singular continuous and absolutely continuous
diffraction; see \cite{BG08,BG09} and references therein. Let us
mention in passing that making a connection with the spectra of
Schr\"odinger operators is tempting, but has so far eluded all
attempts to substantiate it.

In this article, within the realm of pure point diffractive systems, we
consider two examples of limit periodic structures, the period
doubling sequence in one dimension (Section \ref{sec:pd}) and point sets
related to the chair tiling in two dimensions (Section
\ref{sec:chair}). The corresponding point sets arise from fixed points
of (block) substitution rules, but also have a description as cut and
project sets with $2$-adic internal spaces.  We begin with a concise
survey of the relevant quantities, where we restrict ourselves to
measures that are supported on integer lattices. Then, the two main
examples are presented in an informal way, though all results are
rigorous (with the underlying mathematical framework being available
via \cite{Hof,BMS,BM}). A more detailed exposition will appear
in \cite{ao}.

\section{Autocorrelation and diffraction of Dirac combs}

Consider the (weighted) Dirac comb $\omega = \sum_{x\in\mathbb{Z}^{d}}
w(x) \delta_{x}$, with $\delta_{x}$ the normalised point (or Dirac)
measure at $x\in\mathbb{Z}^{d}$ and (in general, complex) weights
$w(x)$. The \emph{natural autocorrelation} (if it exists) of $\omega$
is defined as the Eberlein convolution
\begin{equation} \label{eq:auto}
   \gamma \, = \, \omega \circledast \widetilde{\omega} \,
    := \lim_{N\to\infty} \frac{\omega^{}_{N}  \ast 
        \widetilde{\omega^{}_{N}}}{\mathrm{vol} 
        \bigl( [-N,N]^{d} \bigr)} ,
\end{equation}
which we view as a measure on $\mathbb{R}^{d}$. Here, $\omega^{}_{N} =
\omega |_{[-N,N]^{d}}$ denotes the restriction of $\omega$ to the
closed cube $[-N,N]^{d}$, and $\widetilde{\mu}$ is the `flipped over'
measure, defined by $\widetilde{\mu} (g) = \overline{\mu
  (\widetilde{g})}$ with $g$ a continuous test function of compact
support and $\widetilde{g} (x) = \overline{g (-x) }$, the bar denoting
complex conjugation. In all cases considered below, the limit in
\eqref{eq:auto} exists. Due to $\mathrm{supp} (\omega) \subset
\mathbb{Z}^{d}$, the autocorrelation is then of the form
\begin{equation} \label{eq:auto-form}
   \gamma \, = \sum_{z\in\mathbb{Z}^{d}} \eta(z) \, \delta_{z}
\end{equation}
with the autocorrelation coefficients 
\begin{equation} \label{eq:auto-coeff}
    \eta(z) \,  =  \lim_{N\to\infty} \frac{1}{(2N+1)^d}
        \sum_{x\in \mathbb{Z}^{d}\cap [-N,N]^d}  w(x)\, \overline{w(x-z)} .
\end{equation}
The existence of the limit in \eqref{eq:auto} is equivalent to the
existence of the autocorrelation coefficients $\eta(z)$ for all
$z\in\mathbb{Z}^{d}$.

By construction, $\gamma$ is a positive definite measure, and its
Fourier transform $\widehat{\gamma}$ thus exists. The latter is a
positive, translation bounded measure on $\mathbb{R}^{d}$ (by the
Bochner-Schwartz theorem), which is known as the \emph{diffraction
  measure} of $\omega$. It corresponds to the (kinematic) diffraction
from the measure $\omega$; compare \cite{Cowley} for background and
\cite{Hof,BM} for this approach to diffraction theory. The diffraction
measure possesses the unique decomposition
\begin{equation} \label{eq:decomp}
   \widehat{\gamma} =
   \bigl(\widehat{\gamma}\bigr)_{\mathsf{pp}} +
   \bigl(\widehat{\gamma}\bigr)_{\mathsf{sc}} +
   \bigl(\widehat{\gamma}\bigr)_{\mathsf{ac}}
\end{equation}
into its pure point, singular continuous and absolutely continuous
parts, the latter splitting relative to Lebesgue measure, the natural
reference measure for volume in Euclidean space. Lattice periodic
measures in $\mathbb{R}^{d}$ give rise to pure point diffraction, as
do regular model sets with Euclidean internal space $\mathbb{R}^{m}$,
where the diffraction measure is supported on a $\mathbb{Z}$-module of
finite rank $m+d$; see \cite{BM} and references therein for more.  The
examples discussed below are also based on model sets, but with
$2$-adic internal spaces; they are \emph{limit periodic}, and thus
exhibit pure point diffraction where the corresponding Fourier module
is countably, but not finitely generated. For both examples, the
explicit form of the diffraction can be obtained either from the cut
and project description \cite{BM} or from the inflation structure
\cite{GK}; in this note, we use the latter approach.

\section{The period doubling sequence}
\label{sec:pd}

The \emph{period doubling sequence} is defined by the primitive
substitution rule
\begin{equation} \label{eq:pd-def}
    \varrho : \;\begin{array}{r@{\;}c@{\;}l}
    a & \mapsto & ab \\ b & \mapsto & aa
    \end{array}
\end{equation}    
on the two-letter alphabet $\{a,b\}$. It is related to the classic
Thue-Morse system (as a factor under a $2\! :\! 1$ mapping; compare
\cite{BG08} and references therein) and defines a strictly ergodic
dynamical system via the orbit closure of a suitable bi-infinite
sequence. The latter can be obtained from a fixed point of
$\varrho^{2}$ via the iteration
\begin{equation} \label{eq:pd-iter}
    a | a \xrightarrow{\,\varrho^{2}\,} abaa | abaa 
    \xrightarrow{\,\varrho^{2}\,}
    \ldots \longrightarrow w = \varrho^{2} (w) ,
\end{equation}
with convergence in the (obvious) product topology. Here, $|$ marks
the origin of a two-sided sequence $w = \ldots w^{}_{-2} w^{}_{-1} |
w^{}_{0} w^{}_{1} \ldots$, and we start with the legal seed $a|a$.  The
half-infinite word $v=w^{}_{0} w^{}_{1} \ldots$ is the (unique)
one-sided fixed point of $\varrho$.

To the sequence $w$, we associate two point sets $\varLambda_{\ell}
\subset \mathbb{Z}$ with $\ell\in\{a,b\}$ via
\[
   \varLambda_{\ell} = \{n\in\mathbb{Z}\mid w_{n}=\ell\}  , 
\]
so $\varLambda_{a}\dot{\cup}\varLambda_{b}=\mathbb{Z}$, where
$\dot{\cup}$ denotes the disjoint union of sets.  The geometric fixed
point equation for $\varrho^{2}$ now implies the equations
\begin{subequations}
\begin{align}
   \varLambda_{a} &= 4\varLambda_{a} \,\dot{\cup}\, (4\varLambda_{a}+2) 
              \,\dot{\cup}\, (4\varLambda_{a}+3)
               \,\dot{\cup}\, 4\varLambda_{b} \,\dot{\cup}\, 
                 (4\varLambda_{b}+2),\\
   \varLambda_{b} &= (4\varLambda_{a}+1) \,\dot{\cup}\, 
            (4\varLambda_{b}+1) 
        \,\dot{\cup}\, (4\varLambda_{b}+3),
\end{align}
\end{subequations}
which can be decoupled and solved by iteration, yielding
\begin{equation}\label{eq:pdsol}
   \varLambda_{a} = \{-1\}  \;\dot{\cup}\;\,
    \dot{\bigcup}_{i\ge 0} \bigl(2\!\cdot\! 
     4^i\mathbb{Z} + (4^{i}-1)\bigr), \quad
   \varLambda_{b} = \dot{\bigcup}_{i\ge 1} \bigl(4^i\mathbb{Z} + 
    (2\!\cdot\! 4^{i-1}-1)\bigr),
\end{equation}
where the singleton set $\{-1\}$ has to be added to the fixed point
(to either $\varLambda_{a}$ or $\varLambda_{b}$), as it is only a
limit point under $2$-adic completion \cite{BMS}.

We attach a Dirac comb to $\varLambda_{\ell}$ by $\omega^{}_{\ell} :=
\delta{}_{\varLambda_{\ell}} = \sum_{n\in\varLambda_{\ell}}
\delta^{}_{n}$ for $\ell\in\{a,b\}$. A general weighted Dirac comb
with weights $\alpha$ (for letter $a$) and $\beta$ (for letter $b$)
can then be expressed as $\omega =
\alpha\omega^{}_{a}+\beta\omega^{}_{b}$. Its autocorrelation measure
exists and has the form of Eq.~\eqref{eq:auto-form} (with $d=1$) with
autocorrelation coefficients $\eta(z)$ according to
\eqref{eq:auto-coeff}, where the weight $w(x) \in\{\alpha,\beta\}$ is
chosen according to the letter at position $x \in \mathbb{Z}$.

For simplicity, we work with the one-sided fixed point $v$ (which has
the same autocorrelation). We first consider weights $\alpha=1$ and
$\beta=-1$, and denote the autocorrelation coefficients for this
balanced case by $\eta^{}_{\pm}(z)$. The corresponding sequence
$\{v^{}_{i}\}^{}_{i\in\mathbb{N}_{0}}$ of weights $v_{i}\in\{\pm 1\}$
satisfies the recursions $v^{}_{2n}=1$ and $v^{}_{2n+1}=-v^{}_{n}$ for
$n\ge 0$.  This implies a recursion for the autocorrelation
coefficients,
\begin{equation}
    \eta^{}_{\pm}(2m)= \frac{1}{2}\bigl(1+\eta^{}_{\pm}(m)\bigr) 
    \quad\mbox{and}\quad
    \eta^{}_{\pm}(2m+1)=-\frac{1}{3},
\end{equation}
for all $m\ge 0$. In particular, $\eta^{}_{\pm}(0)=1$, and we also
have $\eta^{}_{\pm}(-m)=\eta^{}_{\pm}(m)$. When $m=2^{r}(2s+1)$ with
$r,s\ge 0$, the recursion leads to 
\[
    \eta^{}_{\pm}(m)=1-\frac{1}{3\cdot 2^{r-2}} ,
\] 
which is independent of $s$. Going back to general weights $\alpha$
and $\beta$ amounts to considering the bi-infinite sequence
$\frac{1}{2}\bigl((\alpha+\beta)\mathbf{1}+ (\alpha-\beta)w\bigr)$, or
alternatively $\frac{1}{2}\bigl((\alpha+\beta)\mathbf{1}+
(\alpha-\beta)v\bigr)$ for its one-sided counterpart, which results in
\begin{equation}
   \eta \,=\, \tfrac{1}{4}(\alpha-\beta)^2 \eta^{}_{\pm} +
   \tfrac{1}{12} \bigl(3(\alpha+\beta)^2+2(\alpha^2-\beta^2)\bigr)
   \mathbf{1} .
\end{equation}
Taking the Fourier transform, the diffraction measure is obtained as
\begin{equation}\label{eq:pd-diff-form}
     \widehat{\gamma}\, =  
     \sum_{k\in L^{\circledast}}
     \bigl| \alpha  A(k) + \beta  B(k) \bigr|^{2} 
     \, \delta^{}_{k}  ,
\end{equation}
where $L^{\circledast} = \bigcup_{s\ge 1} \mathbb{Z}/2^{s} = \bigl\{
\frac{m}{2^r} \mid (r=0 , \, m\in\mathbb{Z}) \mbox{ or } (r\ge 1 , \,
m \mbox{ odd}) \bigr\}$ is the \emph{Fourier module} of the period
doubling sequence.  The amplitudes read
\begin{equation}\label{eq:pdampli}
   A(k) = \frac{2}{3\cdot (-2)^{r}}\,
   e^{2\pi i k}   \quad \mbox{ and } \quad
   B(k) = \delta^{}_{r,0} - A(k) ,
\end{equation}
where we implicitly refer to the parametrisation of $L^{\circledast}$. 

To derive these formulas, one observes that
a finite union in Eq.~\eqref{eq:pdsol} would still be a periodic
set, whose diffraction can be calculated by direct Fourier transform
(which is then a periodic point measure) followed by taking the
absolute square of the amplitudes (note that the Wiener diagram
is still commutative in this case). This results in a sequence of
pure point measures that converge not only in the vague topology,
but also in the stronger norm topology towards the diffraction 
measure in \eqref{eq:pd-diff-form}, with the amplitudes as in
\eqref{eq:pdampli}; see \cite{BM} for details. Since pure point
measures are closed in the norm topology, this also gives the
constructive proof that $L^{\circledast}$ is indeed the Fourier
module.

\begin{figure}
\begin{center}
\includegraphics[width=0.85\textwidth]{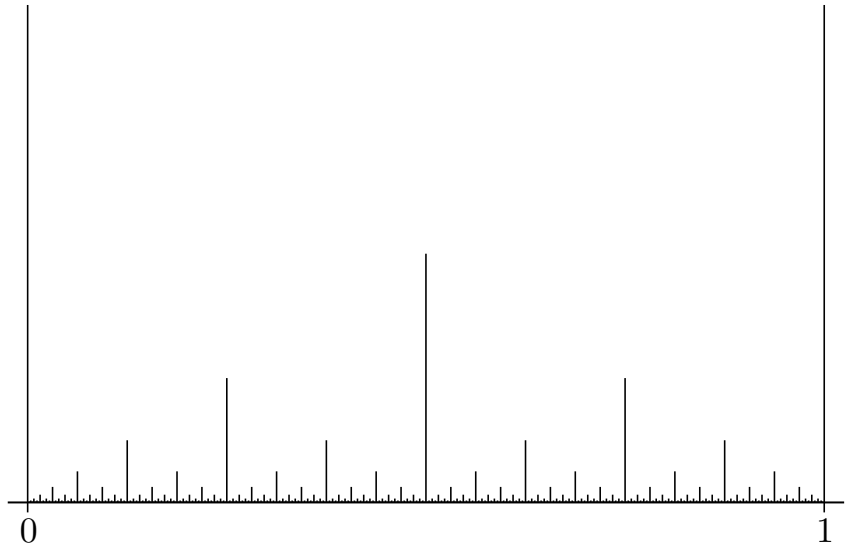}
\caption{Sketch of the diffraction amplitudes $\lvert A(k)\rvert$ from
  Eq.~\eqref{eq:pdampli} for the period doubling chain, with $k\in
  [0,1]$. The diffraction pattern repeats $\mathbb{Z}$-periodically
  along the real line \cite{B}.\label{fig:pddiff}}
\end{center}
\end{figure}

As mentioned above, the same result can also be derived via
the description of the periodic doubling sequence as a model set
with $2$-adic internal space; see \cite{BMS,BM} for a proof. A sketch
of the diffraction pattern is shown in Figure~\ref{fig:pddiff}.

The presence of the powers of $2^{-s}$ in the Fourier module shows the
limit periodic structure; since $s$ can be arbitrarily large, the
Fourier module cannot be finitely generated. Nevertheless, the
diffraction measure is pure point, with amplitudes that reflect the
inflation symmetry with integer multiplier $2$, as visible in
Figure~\ref{fig:pddiff}.

\section{Block substitution for the chair tiling}\label{sec:chair}

Let us move on to our planar example. Consider the block substitution
rule
\begin{equation}\label{eq:blocksub}
   0\,\mapsto\begin{array}{c@{\;\;}c}1&0\\[-1mm] 0&3\end{array}\qquad
   1\,\mapsto\begin{array}{c@{\;\;}c}1&2\\[-1mm] 0&1\end{array}\qquad
   2\,\mapsto\begin{array}{c@{\;\;}c}1&2\\[-1mm] 2&3\end{array}\qquad
   3\,\mapsto\begin{array}{c@{\;\;}c}3&2\\[-1mm] 0&3\end{array}
\end{equation}
on the alphabet $\{0,1,2,3\}$.
Iterating the rule, starting from a legal seed, 
\begin{equation}\label{eq:blockchairfix}
   \begin{array}{c@{\;}|@{\;}c}3&0\\[-0.5mm]\hline 2&1\end{array}
   \;\;\longmapsto\;\:\begin{array}{c@{\;\;}c@{\;}|@{\;}c@{\;\;}c}
    3&2&1&0\\[-1mm]
    0&3&0&3\\[-0.5mm] \hline 
    1&2&1&2\\[-1mm]
    2&3&0&1 \end{array}
 \;\;\longmapsto\;\:\begin{array}{c@{\;\;}c@{\;\;}c@{\;\;}c@{\;}|@{\;}c
   @{\;\;}c@{\;\;}c@{\;\;}c}
    3&2&1&2&1&2&1&0\\[-1mm] 
    0&3&2&3&0&1&0&3\\[-1mm]
    1&0&3&2&1&0&3&2\\[-1mm]
    0&3&0&3&0&3&0&3\\[-0.5mm]   \hline 
    1&2&1&2&1&2&1&2\\[-1mm]
    0&1&2&3&0&1&2&3\\[-1mm]
    1&2&3&2&1&0&1&2\\[-1mm]
    2&3&0&3&0&3&0&1\end{array}
  \;\;\longmapsto\: \; \cdots
\end{equation}
produces a sequence of growing square-shaped blocks (with lines
denoting the coordinate axes, so the origin is located in the centre
of the blocks), where each block reappears in the centre of the
next. This sequence thus converges in the product topology, where the
limit is a fixed point of the block substitution rule that covers the
entire square lattice. This fixed point has a $D_{4}$ colour symmetry:
An anti-clockwise rotation through $\pi/2$ corresponds to the cyclic
permutation $(0123)$, while a reflection in the central horizontal
gives the permutation $(01)(23)$. The pattern is thus invariant under
any $D_{4}$ transformation followed by an appropriate permutation.

If we choose a representation with four different unit squares and
take the lower left corners as their reference points, the fixed point
defined by \eqref{eq:blockchairfix} results in a partition
$\mathbb{Z}^{2} = \varLambda^{}_{0} \dot{\cup} \varLambda^{}_{1}
\dot{\cup} \varLambda^{}_{2}\dot{\cup} \varLambda^{}_{3}$ of
$\mathbb{Z}^{2}$ into four point sets of equal density
$\frac{1}{4}$. As a result of \eqref{eq:blocksub}, they satisfy the
fixed point equations
\begin{subequations}
\begin{align}
   \varLambda^{}_{0} &= 2\varLambda^{}_{0}\,\dot{\cup}\; 
    (2\varLambda^{}_{0}+u+v) 
                \,\dot{\cup}\, 2\varLambda^{}_{1} 
                \,\dot{\cup}\, 2\varLambda^{}_{3} , \\
   \varLambda^{}_{1} &= (2\varLambda^{}_{0}+v) \,\dot{\cup}\; 
    (2\varLambda^{}_{1}+u) 
                \,\dot{\cup}\; (2\varLambda^{}_{1}+v) 
                \,\dot{\cup}\; (2\varLambda^{}_{2}+v) , \\
   \varLambda^{}_{2} &= (2\varLambda^{}_{1}+u+v)\,\dot{\cup}\, 
    2\varLambda^{}_{2} 
                \,\dot{\cup}\; (2\varLambda^{}_{2}+u+v) 
                \,\dot{\cup}\; (2\varLambda^{}_{3}+u+v) , \\
   \varLambda^{}_{3} &= (2\varLambda^{}_{0}+u) \,\dot{\cup}\; 
    (2\varLambda^{}_{2}+u) 
                \,\dot{\cup}\; (2\varLambda^{}_{3}+u) 
                \,\dot{\cup}\; (2\varLambda^{}_{3}+v) ,
\end{align}
\end{subequations}
with $u=(1,0)^{T}$ and $v=(0,1)^{T}$. Denoting the even and odd
sublattices of $\mathbb{Z}^{2}$ as
$\varGamma_{+}=\bigl\{(x_{1},x_{2})^{T}\in\mathbb{Z}^{2}\;\big|\;
x_{1}+x_{2}\equiv 0 \bmod 2\bigr\}$ and
$\varGamma_{-}=u + \varGamma_{+}$, we have
$\mathbb{Z}^{2}=\varGamma_{+}\dot{\cup}\varGamma_{-}$, as well as
$\varLambda^{}_{0}\dot{\cup}\varLambda^{}_{2} = \varGamma^{}_{+}$ and
$\varLambda^{}_{1}\dot{\cup}\varLambda^{}_{3} = \varGamma^{}_{-}$.
This leads to the decoupled equations
\begin{equation}
   \varLambda^{}_{0} = 2\varGamma^{}_{-} \,\dot{\cup}\; 
   \bigl(2\varLambda^{}_{0}+\{0,u+v\}\bigr)
    \quad\text{and}\quad
   \varLambda^{}_{1} = \bigl(2\varGamma^{}_{+}+v\bigr) \,\dot{\cup}\; 
   \bigl(2\varLambda^{}_{1}+\{u,v\}\bigr), 
\end{equation}
together with
$\varLambda^{}_{2}=\varGamma^{}_{+}\setminus\varLambda^{}_{0}$ and
$\varLambda^{}_{3}=\varGamma^{}_{-}\setminus\varLambda^{}_{1}$. An
explicit solution is (with $Sx:=\{sx\mid s\in
S\}$ for sets $S\subset\mathbb{Z}$ and
$S_{r}:=\{0,1,\ldots,2^{r}\!-1\}$ for $r\ge 0$)
\begin{subequations}\label{eq:chairsol} 
   \begin{align}
   \varLambda^{}_{0} &\,= \hphantom{-}\mathbb{N}_{0}^{} (u+v) \;\dot{\cup}\;
                \dot{\bigcup}_{r\ge 0}\, 
                \bigl( 2^{r+1} \varGamma^{}_{-} + 
                S_{r} (u+v)\bigr), \\
   \varLambda^{}_{1}+v &\,= \hphantom{-}\mathbb{N}_{0}^{} (u-v) \;\dot{\cup}\;
                \dot{\bigcup}_{r\ge 0}\, 
                \bigl( 2^{r+1} \varGamma^{}_{-} +
                S_{r} (u-v)\bigr),\\
   \varLambda^{}_{2}+u+v &\,= -\mathbb{N}_{0}^{} (u+v) \;\dot{\cup}\;
                \dot{\bigcup}_{r\ge 0}\, 
                \bigl( 2^{r+1} \varGamma^{}_{-} -
                S_{r} (u+v)\bigr),\\
   \varLambda^{}_{3}+u &\,= -\mathbb{N}_{0}^{} (u-v) \;\dot{\cup}\;
                \dot{\bigcup}_{r\ge 0}\, 
                \bigl( 2^{r+1} \varGamma^{}_{-} -
                S_{r} (u-v)\bigr),
    \end{align}
\end{subequations}
where we used
$\varGamma_{+}+u=\varGamma_{+}+v=\varGamma_{-}=\varGamma_{-}+u+v$. Each
of the infinite unions on the right-hand side defines a point set of
density $\frac{1}{4}$. The sets of points (of density $0$) along the
diagonals have to be added for our fixed point
\eqref{eq:blockchairfix}, because they are not contained in any of the
lattice translates (but are covered by the $2$-adic completion of the
infinite unions, similar to the situation of the period doubling
sequence). The solutions obey the relations
$\varLambda^{}_{0}\dot{\cup}\varLambda^{}_{2}=\varGamma^{}_{+}$ and
$\varLambda^{}_{1}\dot{\cup}\varLambda^{}_{3}=\varGamma^{}_{-}$.  

Now, we are going to calculate the diffraction measure of the Dirac comb
\begin{equation}\label{eq:chaircomb}
    \omega = \sum_{i=0}^{3}\alpha^{}_{i}\,\delta^{}_{\varLambda_{i}}
\end{equation}
with the four point sets $\varLambda^{}_{i}$ from
Eq.~\eqref{eq:chairsol}, and arbitrary complex numbers $\alpha_{i}$.
 
Define $Q_{r}(x):=2^{r+1} \varGamma_{-} + S_{r} x$ for
$r\in\mathbb{N}_{0}^{}$ and $x\in\mathbb{Z}^{2}$.  Using
$\varGamma_{+}^{*}=\frac{1}{2}\varGamma^{}_{+}$ (where the superscript
$*$ denotes the dual lattice), we calculate
\begin{align}
     \widehat{\delta^{}_{Q_{r}(x)}} \:&=\,
     \sum_{m=0}^{2^{r}\!-1} \mathrm{e}^{-2\pi\mathrm{i} m kx}\,
     \bigl(\delta^{}_{2^{r+1}(\varGamma_{+}+u)}\bigr)\!\widehat{\phantom{T}} 
     =\, \frac{1-\mathrm{e}^{-2^{r+1}\pi\mathrm{i} kx}}
      {1-\mathrm{e}^{-2\pi\mathrm{i} kx}}\,
       \mathrm{e}^{-2^{r+2}\pi\mathrm{i} ku} \widehat{\,
      \delta^{}_{2^{r+1}\varGamma_{+}}}\notag\\
     &= \,\frac{1-\mathrm{e}^{-2^{r+1}\pi\mathrm{i} kx}}
      {1-\mathrm{e}^{-2\pi\mathrm{i} kx}}\,
       \frac{\mathrm{e}^{-2^{r+2}\pi\mathrm{i} ku}}{2^{2r+3}_{}}\,
       \delta^{}_{\varGamma_{+}/2^{r+2}} ,\label{eq:amplisum}
\end{align}
where we again write density factors as functions of $k$. 
The first fraction evaluates as $2^r$ whenever $kx\in\mathbb{Z}$
(by an application of the l'Hospital rule). 
The resulting Fourier module is
\begin{equation}\label{eq:block-module}
      L^{\circledast}  =\, \bigcup_{s \ge 2} \frac{\varGamma_{+}}{2^{s}}
           \:= \:\mathbb{Z}^{2} \;\dot{\cup}\; \dot{\bigcup}_{s\ge 1}
             \Bigl\{ \frac{1}{2^{s}} (m,n) \mid m,n\in\mathbb{Z} 
             \text{ with } \gcd(m,n,2)=1 \Bigr\},
\end{equation}
which is the coarsest group that contains all positions where the
Fourier transform in \eqref{eq:amplisum} has a point measure.

The diffraction measure of our weighted Dirac comb $\omega$ is of the form
\begin{equation}\label{eq:wdc}
   \widehat{\gamma^{}_{\omega}}\: = \sum_{k\in L^{\circledast}}\,
    \Bigl| \sum_{i=0}^{3} \alpha_{i} \, A_{i}(k) \Bigr|^{2} \delta_{k} ,
\end{equation}
where $A_{i}(k)$ is the amplitude that belongs to $\varLambda_{i}$.
Its derivation is analogous to the method explained above for the
period doubling sequence (via periodic approximants that converge
in the norm topology). This method also provides a proof for the
correctness of \eqref{eq:block-module}.

The amplitudes themselves can be calculated in the same way as for the
period doubling sequence above.  For any $k\in L^{\circledast}$, there
is a minimal $s\ge 0$ such that $k\in\varGamma_{+}/2^{s+2}$, and then
$k\in\varGamma_{+}/2^{r+2}$ for all $r\ge s$. The amplitudes result
from summing the corresponding contributions from
Eq.~\eqref{eq:amplisum}, with $x=u+v$ for $i=0$, $x=u-v$ for $i=1$,
$x=-u-v$ for $i=2$ and $x=-u+v$ for $i=3$. In addition, the results
have to be multiplied by an overall phase factor that reflects the
translation vectors on the left-hand sides of the four relations in
Eq.~\eqref{eq:chairsol}. We omit the detailed (and somewhat lengthy)
calculations and simply state the result.

To this end, the integer and half-integer points of $L^{\circledast}$
are better treated separately. Whenever $k\in\mathbb{Z}^{2}$,
\begin{equation}
    A_{0}(k)=A_{1}(k)=A_{2}(k)=A_{3}(k)=\frac{1}{4} .
\end{equation}
If $k\in\frac{1}{2}\varGamma_{+}\setminus \mathbb{Z}^{2}$, one has
$k=\frac{1}{2}(m,n)$ with $m$ and $n$ odd, so that
$kx\in\mathbb{Z}$ for all $x\in\varGamma_{+}$. This gives
\begin{equation}
    A_{0}(k)=A_{2}(k)=\frac{1}{4}\quad \text{and}\quad
    A_{1}(k)=A_{3}(k)=-\frac{1}{4} ,
\end{equation}
while $k=\frac{1}{2}(m,n)$ with $m+n$ odd results in
\begin{equation}
    A_{0}(k)=\frac{1}{8} , \quad
    A_{2}(k)=-\frac{1}{8} , \quad
    A_{1}(k)=\frac{(-1)^{n}}{8} \quad \text{and}\quad
    A_{3}(k)=\frac{(-1)^{m}}{8} .
\end{equation}
In particular, $A_{0}(k)+A_{2}(k)=0$, and also $A_{1}(k)+A_{3}(k)=0$,
as $m+n$ is odd.

The remaining elements of $L^{\circledast}$ are of the form
$k=(m,n)/2^{s}$ with $s\ge 2$ and $m,n\in\mathbb{Z}$ subject to the
condition $\gcd(m,n,2)=1$. The amplitudes $A_{0}(k)$ and $A_{2}(k)$
depend on $m+n$. They satisfy $A_{2}(k)=-A_{0}(k)$ and
\begin{equation}
    A_{0}(k)\, = \,\begin{cases}
    0 , & \text{if $2^{s}\mid (m+n)$}, \\
    \displaystyle -\frac{2}{4^{s}}\,
    \frac{1-(-1)^{(m+n)/2}}{1-\varepsilon_{s}^{m+n}},
    & \text{if $m+n$ even with $2^{s}\nmid (m+n)$},\\
    \displaystyle\frac{1}{4^{s}}\,\frac{1}{1-\varepsilon_{s}^{m+n}},
    & \text{if $m+n$ odd},
               \end{cases}
\end{equation}
where $\varepsilon^{}_{s}=\exp(-2\pi\mathrm{i}/2^{s})$ is a root of
unity. Similarly, one obtains the relation $A_{3}(k)=-A_{1}(k)$
together with
\begin{equation}
    A_{1}(k)\, =\, \varepsilon_{s}^{-n}\begin{cases}
    0 , & \text{if $2^{s}\mid (m-n)$}, \\ \displaystyle
    -\frac{2}{4^{s}}\,\frac{1-(-1)^{(m-n)/2}}
      {1-\varepsilon_{s}^{m-n}},
    & \text{if $m-n$ even with $2^{s}\nmid (m-n)$},\\
    \displaystyle \frac{1}{4^{s}}\,\frac{1}{1-\varepsilon_{s}^{m-n}},
    & \text{if $m-n$ odd}.
               \end{cases}
\end{equation}
In particular, 
$\alpha^{}_{0}=\alpha^{}_{1}=\alpha^{}_{2}=\alpha^{}_{3}=1$ leads to
$\widehat{\gamma^{}_{\omega}}=\delta_{\mathbb{Z}^{2}}^{}$, as
required. Moreover, setting $\alpha^{}_{0}=\alpha^{}_{2}=1$ and
$\alpha^{}_{1}=\alpha^{}_{3}=0$ gives the diffraction
$\widehat{\gamma^{}_{\omega}}=\widehat{\delta^{}_{\varGamma_{+}}\!}=
\frac{1}{2}\delta^{}_{\varGamma_{+}/2}$, while the alternative choice
$\alpha^{}_{0}=\alpha^{}_{2}=0$ together with
$\alpha^{}_{1}=\alpha^{}_{3}=1$ yields (using
the parametrisation of $k$ as above)
$\widehat{\gamma^{}_{\omega}}=\widehat{\delta^{}_{\varGamma_{-}}\!}=
\mathrm{e}^{-2\pi\mathrm{i} ku}\,\widehat{\delta^{}_{\varGamma_{+}}\!}=
\frac{(-1)^{m}}{2}\,\delta^{}_{\varGamma_{+}/2}$.

\begin{figure}
\begin{center}
\includegraphics[width=0.8\textwidth]{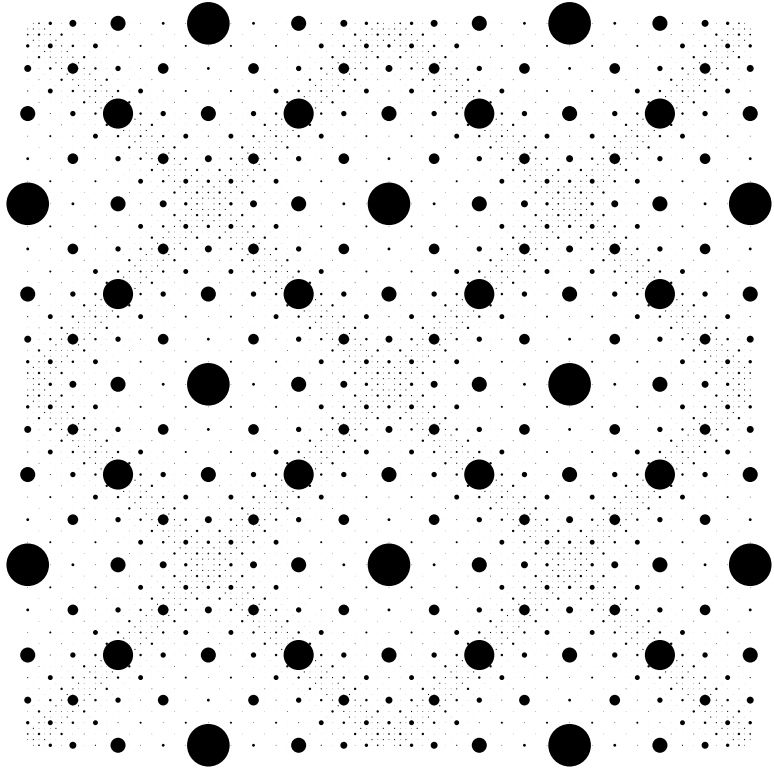}
\caption{Illustration of the diffraction measure for the Dirac comb
  \eqref{eq:wdc} with weights $\alpha_{j}=\mathrm{i}^{j}$, where
  $j\in\{0,1,2,3\}$. A peak at $k\in [-1,1]^{2}$ is represented by a black
  disc, centred at $k$, with an area proportional to the peak
  intensity.  The complete pattern is
  $\frac{1}{2}\varGamma_{+}$-periodic.\label{fig:chairdiff}}
\end{center}
\end{figure}
                                                                     
An interesting case results from the balanced choice
$\alpha^{}_{j}=\mathrm{i}^{j}$, which produces extinctions for all
$k\in\frac{1}{2}\varGamma_{+}$. The corresponding diffraction measure
is illustrated in Figure~\ref{fig:chairdiff}. It is $D_4$-symmetric,
which is not the case in general.
                                                                               
For a generic choice of the weights, our Dirac comb
\eqref{eq:chaircomb} is mutually locally derivable (MLD) with the chair
tiling.  As it is based on four subsets of $\mathbb{Z}^{2}$, the
diffraction measure $\widehat{\gamma^{}_{\omega}}$ is always
$\mathbb{Z}^{2}$-periodic \cite{B}. This is the generic maximal translation
symmetry. However, since
$\varLambda_{0}\dot{\cup}\varLambda_{2}=\varGamma_{+}$, the
diffraction of any weighted Dirac comb with only these two components 
is periodic under the dual lattice
$\varGamma_{+}^{*}=\frac{1}{2}\varGamma^{}_{+}$. Likewise, the
corresponding statement holds for the diffraction of weighted Dirac
combs with components $\varLambda_{1}$ and $\varLambda_{3}$. This is
also the translation symmetry in Figure~\ref{fig:chairdiff}, which is
a consequence of the particular choice of the weights here.

Any choice of the weights refers to a discrete structure that can be
viewed as a decoration of the chair tiling, and is thus locally
derivable from it. Conversely, a typical decoration of the chair
tiling will be locally derivable from our coloured point set, but need
not be realisable as a simple decoration of the square-shaped building
blocks. In this sense, our above results constitute only a first step
of the complete diffraction analysis of the chair tiling.

\section{Summary and Outlook}

Limit periodic structures form an interesting class of pure point
diffractive systems. Explicit results for the diffraction can be
obtained, at least for examples such as those discussed in this paper;
related examples can be found in \cite{AMF,Q}.

Limit periodic tilings also arise in the context of Wang
tiles. Raphael Robinson's tilings consisting of six square-shaped
prototiles \cite{Rob71}, up to rotations and reflections, is one
example. In this case, the limit periodic structure is apparent from a
hierarchical pattern of interlocking squares of larger and larger
sizes.  Recently, a similar tiling with a single hexagonal prototile,
up to rotations and reflections, was discovered \cite{ST,T}, with a
hierarchical pattern of interlocking triangles. It is related to the
half-hex inflation that was studied in detail in \cite{F}, where also
its limit-periodic structure was derived, and to Penrose's
$(1+\varepsilon+\varepsilon^2)$ tiling \cite{Pen97}.

The new example is particularly intriguing because the hexagon is the
first example of a single aperiodic tile (provided you count the tile
and its mirror image as a single tile), in the sense that there exist
local rules that allow the existence of tilings covering the entire
plane, but no such tiling can have any non-trivial period. The local
rules involve not just neighbours, but also the next corona, and thus
cannot be encapsulated in the shape of a tile whose interior is
connected; see \cite{ST} for details.  It would be interesting to
study the diffraction properties of these tilings, which will reflect
the limit periodic structure apparent in the patterns of squares and
triangles.

\section*{Acknowledgements}

This work was supported by the German Research Council (DFG), within
the CRC 701, and by a Leverhulme Trust Visiting Professorship Grant (MB).

\end{document}